\documentclass[11pt]{article}
\usepackage{moriond,epsfig,wrapfig}

\def\be{\begin{equation}}
\def\ee{\end{equation}}
\def\bea{\begin{eqnarray}}
\def\eea{\end{eqnarray}}

\begin{document}
\vspace*{4cm}
\title{STATUS OF ICARUS}


\author{J. RICO for the ICARUS Collaboration \footnote{Beijing, CERN,
ETH Z\"urich, Granada, Katowice, Krak\'ow, LNGS, L'Aquila, Milano,
Padova, Pavia, Pisa, Torino, UCLA, Warsawa, Wroclaw}}

\address{Institut f\"ur Teilchenphysik, ETHZ\\ Z\"urich, Switzerland}

\maketitle\abstracts{
The ICARUS detector is a liquid argon time projection chamber. It
provides three dimensional imaging and calorimetry of ionizing
particles over a large volume, with high granularity. Its Physics
program includes the study of atmospheric, solar, supernovae and beam
neutrinos as well as proton decay searches.\\ The ICARUS technology
has reached maturity with the construction and test (during summer
2001) of a 600 ton detector, demonstrating the feasibility of building
large mass devices relevant for non-accelerator physics. During this
test run, more than 27000 cosmic ray events have been acquired. These
data allow to assess the detector performance, i.e.\ the spatial
reconstruction, calorimetry and particle identification.}

\section{Introduction}  

The ICARUS technology, first proposed by C. Rubbia \cite{CARLO} in
1977, combines the characteristics of a bubble chamber with the
advantages of electronic read-out. The detector is an ideal device to
study particle interactions: it is continuously sensitive,
self-triggering, cost effective, simple to build in modular form and
sufficiently safe to be located underground (no pressure, no flammable
gas, etc.). This detector is also a superb calorimeter of very fine
granularity and high accuracy.

A number of test devices of increasing dimensions have been
successfully operated over the years. The latest step on this graded
path, the ICARUS T600 detector, has been fully tested on surface
conditions during 2001. The technical aspects of the system have been
checked during this run, and the acquired data is being used to assess
the spatial and calorimetric reconstruction capabilities of the
detector. In this paper we briefly summarize the results from the T600
technical run and the status of data reconstruction.

\section{The ICARUS T600 detector}   

\subsection{Detector description}

\begin{wrapfigure}{r}{7.5cm}
\vspace{-0.7cm}
\hspace{0.7cm}
\begin{minipage}[c]{6.5cm}
\begin{center}
\epsfig{file=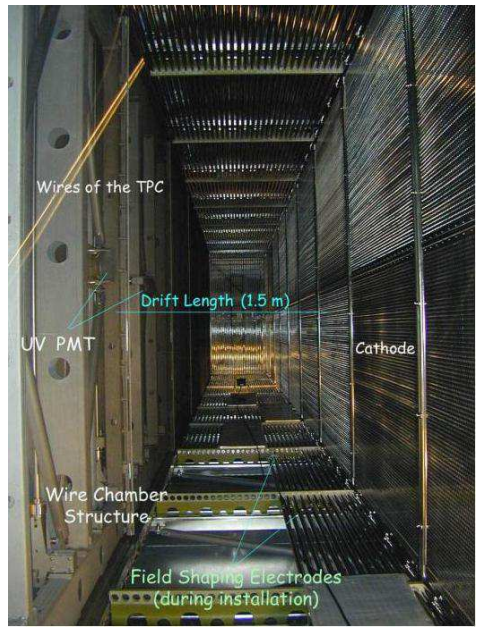,width=6.2cm}
\caption{Internal view of the T600 first half-module}
\label{fig:T600_int}
\end{center}
\end{minipage}
\vspace{-0.5cm}
\end{wrapfigure}

ICARUS T600 \cite{T600} is a large cryostat divided in two identical,
adjacent half-modules of internal dimensions $3.6 \times 3.9 \times
19.9$ m$^3$ each, containing more than 300~t of liquid argon
(LAr). Each half-module houses an internal detector (composed by two
Time Projection Chambers --TPC--, the field shaping system, monitors,
probes, PMT's) and is externally surrounded by a set of thermal
insulations layers. Each TPC is formed by three parallel planes of
wires, 3~mm apart, oriented at $0, \pm60^\circ$ angles, of 3~mm pitch
parallel wires, positioned onto the longest walls of the half-module
(see figure~\ref{fig:T600_int}). A high voltage system produces a
uniform electric field, perpendicular to the wire planes, allowing the
drift of the ionization electrons (the maximum drift path is 1.5~m).

\vspace{-0.2cm}
\subsection{Physics program}
\vspace{-0.2cm}
The initial physics program with the T600 module at Gran Sasso has
been reported elsewhere \cite{REDBOOK}. In this phase the available mass is
limited, however the high efficiency and the detailed information
which can be collected for each event will allow to initiate the study
of some of the fundamental issues of underground physics: the study of
neutrino physics, with solar, atmospheric and supernova neutrinos, and
the study of nucleon decay.

\vspace{-0.2cm}
\subsection{The test run in Pavia}
\vspace{-0.2cm}

\begin{wrapfigure}{r}{7.5cm}
\vspace{-0.7cm}
\hspace{0.7cm}
\begin{minipage}[c]{6.5cm}
\begin{center}
\epsfig{file=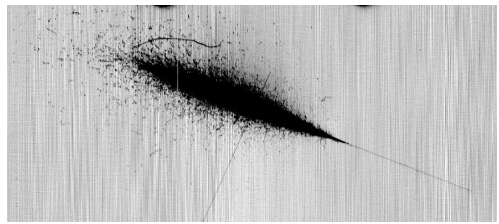,width=6.5cm}
\caption{Electromagnetic shower event from the ICARUS T600 first
half-module test. Event real dimensions: $3.7 \times 1.7$~m$^2$}
\label{fig:shower}
\end{center}
\end{minipage}
\end{wrapfigure}

A full above-ground test of the T600 experimental set-up has been
carried out in Pavia (Italy) during the period April-August 2001. One
T600 half-module has been fully instrumented to allow a complete test
in real experimental conditions. All technical aspects of the system,
cryogenics, mechanics, LAr purification, read-out chambers,
scintillation light detection, electronics, and DAQ have been tested,
and found to be satisfactorily in agreement with expectations.

During the test run, a very large amount of cosmic ray events has been
recorded with different configurations of a dedicated trigger system. A
systematic visual event scanning is being carried out in order to build
an inventory of the acquired data. The current status is shown in
Table~\ref{table:scanning}. An example of a large electromagnetic
shower event acquired during the T600 test is shown in
figure~\ref{fig:shower}.

\begin{table}[t]
\caption{Results from the ICARUS T600 test run inventory. Event
categories are not exclusive.}
\vspace{0.2cm}
\begin{center}
\begin{tabular}{|c|c|}\hline
Event category &  Number of events \\ \hline
Shower & 651 \\ \hline
Muon decay/stopping & 1935 \\ \hline
Hadron interaction & 704 \\ \hline
V$_0$ & 46 \\ \hline
Long track & 311 \\ \hline
Muon bremsstrahlung & 1339 \\ \hline
Multiple showers & 695 \\ \hline
Multiple muons & 138 \\ \hline
Total scanned & 3098 \\ \hline
\end{tabular}
\label{table:scanning}
\end{center}
\vspace{-0.3cm}
\end{table}

\section{ICARUS data reconstruction} 
Like a bubble chamber, the ICARUS detector provides a measurement of
the total ionization loss of a track with very high sampling. Charged
particles traversing the LAr sensitive volume produce ionization
electrons in a number proportional to the energy transferred from the
particle to the LAr. The ionization electrons drift perpendicularly to
the wire planes pushed by the electric field, inducing a signal
(\emph{hit}) on the neighbor wires while approaching the different
wire planes. By extracting the physical information contained in the
wires output signal, i.e.\ the energy deposited by the different
particles and the point where such a deposition has occurred, it is
possible to build a complete three dimensional spatial and
calorimetric picture of the event.

\vspace{-0.2cm}
\subsection{Three dimensional reconstruction}
\begin{figure}
\begin{center}
\parbox{7.5cm}{
\begin{center}
\epsfig{file=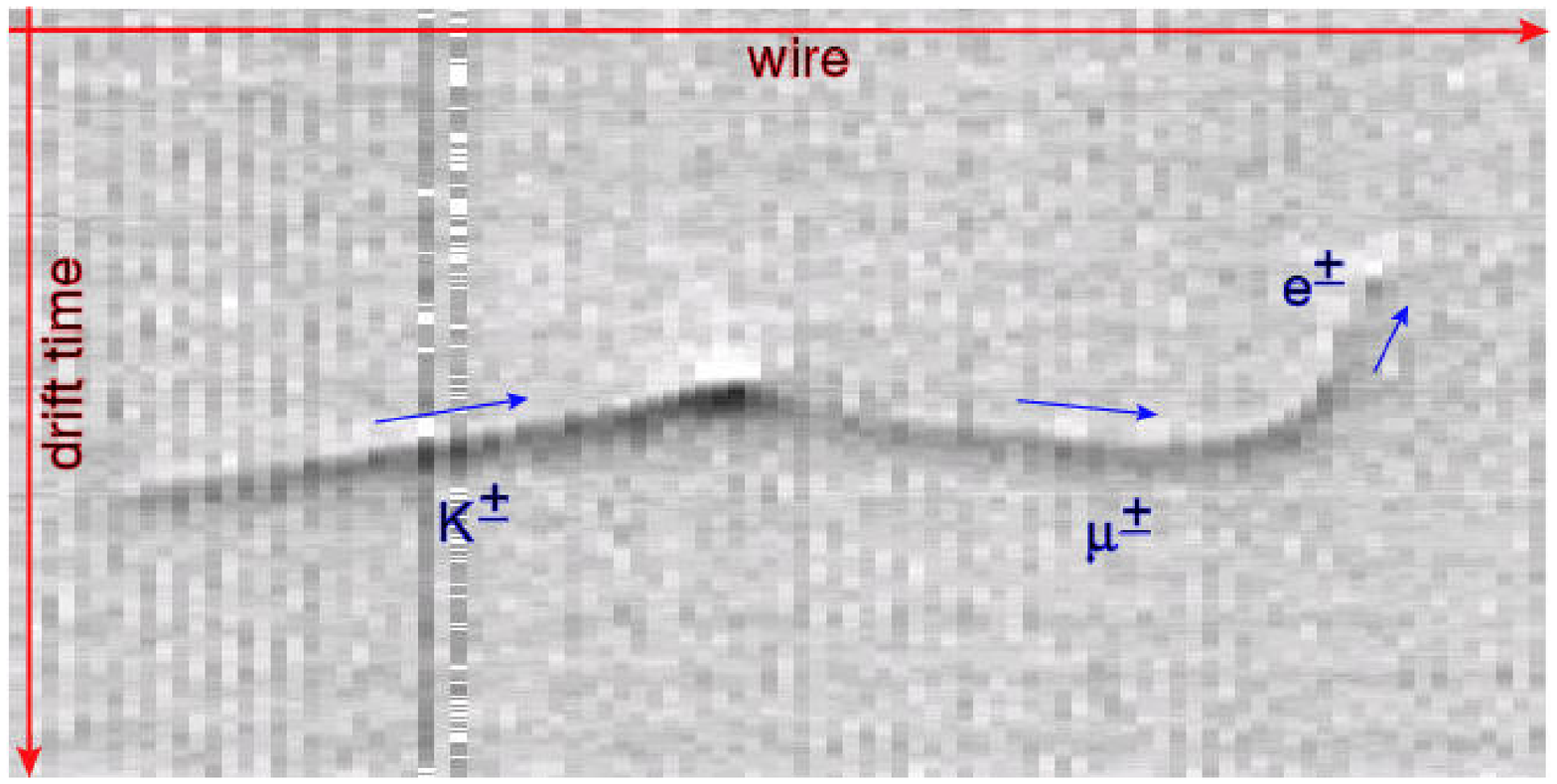,width=6.5cm}
\epsfig{file=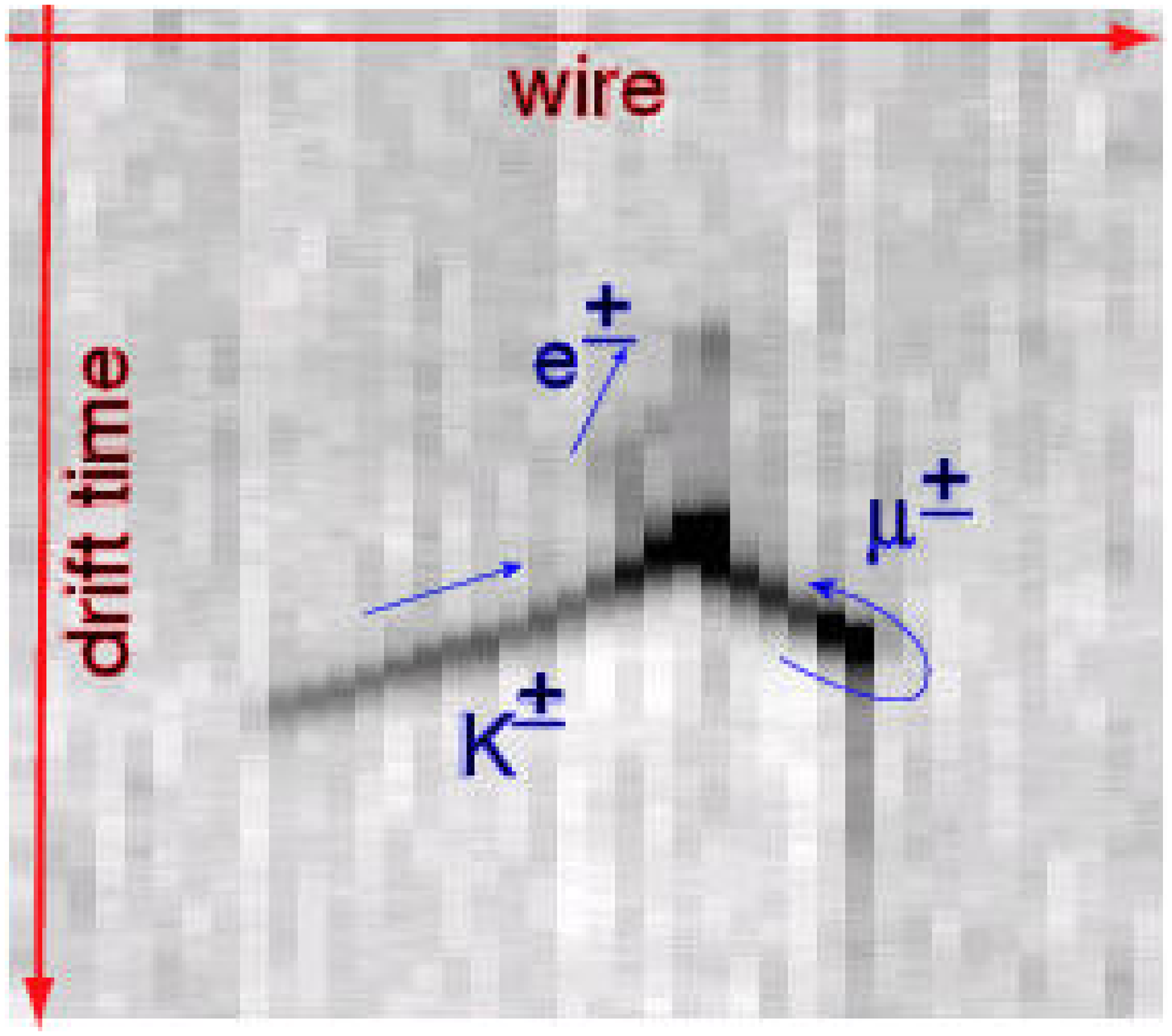,height=3cm}
\epsfig{file=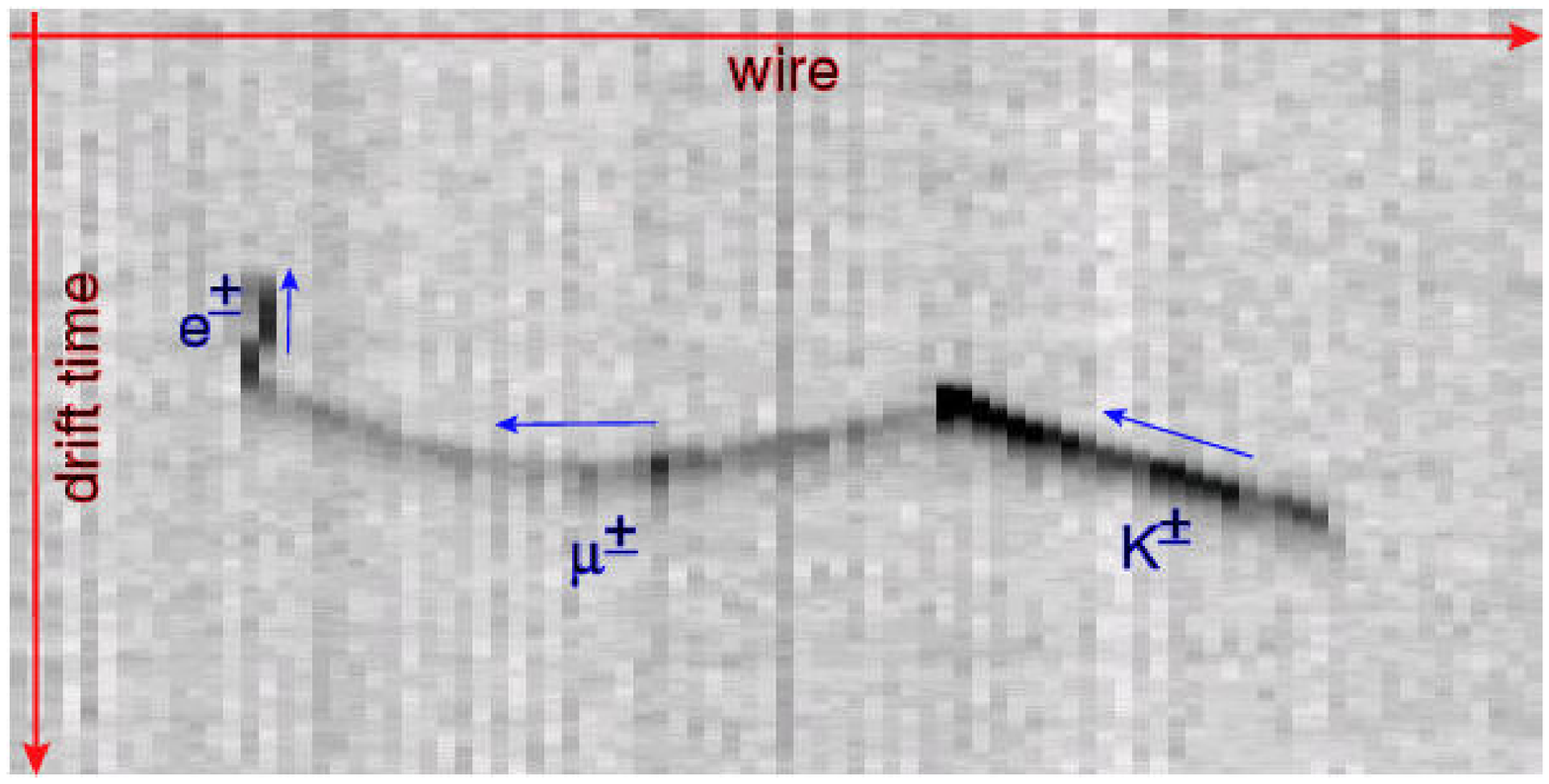,width=6.5cm}
\end{center}
}
\parbox{7.5cm}{
\epsfig{file=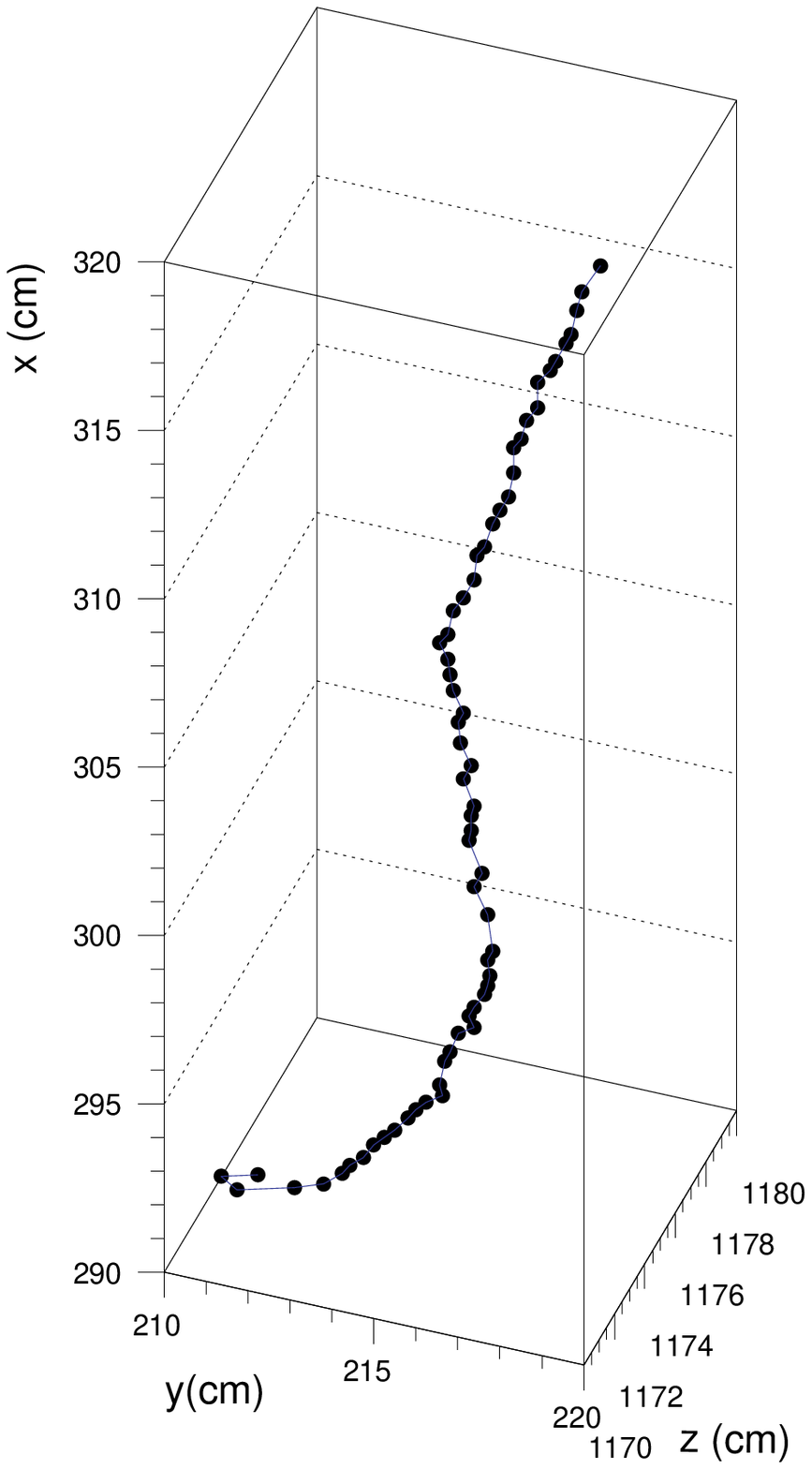,width=6.cm}
}
\end{center}
\caption{$K^\pm$ decay candidate from the T600 half-module test
run. Left: Gray scale representation of the output signal in the wire
(horizontal) vs drift time (vertical) plane for the three wire
planes. Right: Three dimensional reconstruction of the event.}
\label{fig:kaon_3d}
\end{figure}

\vspace{-0.2cm}
Each wire plane constrains two spatial degrees of freedom of the hits,
one common to all the wire planes (the drift time) and one specific
for each plane (the wire coordinate). The redundancy on the drift time
coordinate allows the association of hits from different planes to a
common charge deposition, and together with the wire coordinates from
at least two planes, allow the spatial reconstruction of ionizing
tracks. Figure~\ref{fig:kaon_3d} shows a kaon decay candidate event acquired
during the T600 test. The identification and association of hits found in
the wire output signal of the different wire planes (left)
allow the three dimensional reconstruction of the event (right).

\vspace{-0.2cm}
\subsection{Particle identification}
\begin{figure}
\begin{center}
\epsfig{file=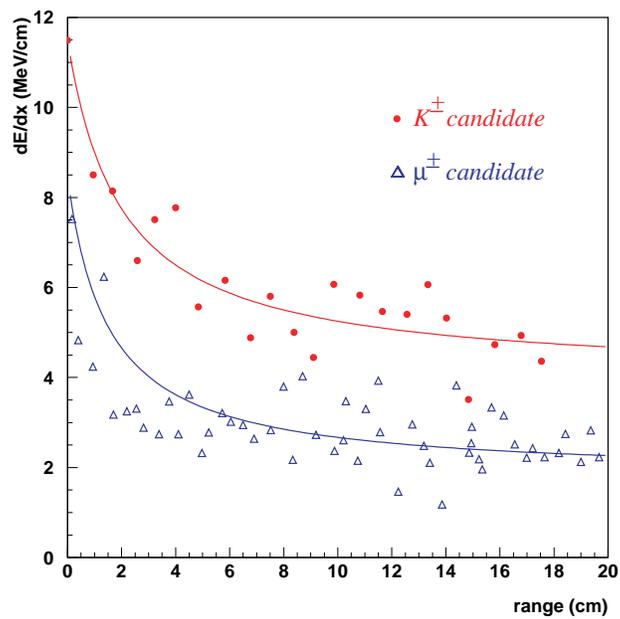,width=9.cm}
\end{center}
\caption{Measured $dE/dx$ vs range for $K^\pm$ and $\mu^\pm$ candidates in
figure~\ref{fig:kaon_3d}.}
\label{fig:kaondEdx}
\end{figure}

\vspace{-0.2cm}
The energy released by ionizing particles per unit length ($dE/dx$)
is, for a given medium, a function of the particle type and its
momentum. In the ICARUS detector, a $dE/dx$ measurement is
performed for a large number of points along a given track. Particle
momentum can be measured from range (for stopping particles) or
multiple scattering measurement, providing a method for particle
identification. Figure~\ref{fig:kaondEdx} shows the measured $dE/dx$
vs range for the $K^\pm$ and $\mu^\pm$ candidates in
figure~\ref{fig:kaon_3d}, proving the discrimination power of this
kind of detector.

\vspace{-0.1cm}
\section{Conclusions} 
\vspace{-0.2cm}
The ICARUS T600 has been successfully tested, demonstrating the
maturity of the ICARUS technology. The acquired data is been used to
assess the performance of the reconstruction tools. The data taking
start up in underground conditions at LNGS is foreseen for the beginning
of 2003.

\vspace{-0.1cm}
\section*{References}     

\end{document}